\documentclass[journal]{IEEEtran}
\usepackage{stmaryrd}
\usepackage{amsmath}
\ifCLASSINFOpdf
\else
\fi
\begin{document}
%
\title{Hard Fault Analysis of Trivium}
%
%
%

\author{Yupu~Hu,~\IEEEmembership{}
        Fengrong~Zhang,~\IEEEmembership{}
        and~Yiwei~Zhang,~\IEEEmembership{}
\thanks{Manuscript received July 1, 2009; This work was supported in part
 by National Science Foundation of China under grant 60833008 and by 973
Project under grant 2007CB311201.}
\thanks{Y.~Hu and F.~Zhang are with the CNIS Laboratory, Xidian University, 710071
Xi'an, China e-mail: (yphu@mail.xidian.edu.cn; zhfl203@163.com).}
\thanks{Y.~Zhang is with ZTE IC Design CO., LTD., 518057 Shenzhen, China
e-mail: (zhang.yiwei2@zte.com.cn).}}

\maketitle

\begin{abstract}
  Fault analysis is a powerful attack to stream ciphers. Up to now,
the major idea of fault analysis is to simplify the cipher system by
injecting some soft faults. We call it soft fault analysis. As a
hardware--oriented stream cipher, Trivium is weak under soft fault
analysis.

   In this paper we consider another type of fault analysis of stream
cipher, which is to simplify the cipher system by injecting some
hard faults. We call it hard fault analysis. We present the
following results about such attack to Trivium. In Case 1 with the
probability not smaller than 0.2396, the attacker can obtain 69 bits
of 80--bits--key. In Case 2 with the probability not smaller than
0.2291, the attacker can obtain all of 80--bits--key. In Case 3 with
the probability not smaller than 0.2291, the attacker can partially
solve the key. In Case 4 with non--neglectable probability, the
attacker can obtain a simplified cipher, with smaller number of
state bits and slower non--linearization procedure. In Case 5 with
non--neglectable probability, the attacker can obtain another
simplified cipher. Besides, these 5 cases can be checked out by
observing the key--stream.
\end{abstract}

\begin{IEEEkeywords}
  Side--channel analysis, fault analysis, stream cipher, Trivium
\end{IEEEkeywords}

%
\IEEEpeerreviewmaketitle


\section{Introduction}
%
%
%
%
%


\subsection{Background and Results of Our Work}
  Side--channel analysis of stream ciphers \cite{C2004} is a class of novel
attacks by combining physical and mathematical methods, including
fault analysis \cite{J2004}, power analysis \cite{W2007}, timing
analysis, etc. In the class of side--channel analysis, fault
analysis is a powerful attack. Up to now, the major idea of fault
analysis is to simplify the cipher system by injecting some soft
faults (that is, by changing the values of some positions at some
moment), thus revealing the key hidden in the encryption machine. We
call such attack soft fault analysis. Soft fault analysis is a known
differential attack \cite{E2007}, by which the attacker can obtain
additional low--degree--equations of the state. Trivium
\cite{C2005,CD} is a hardware--oriented stream cipher, and one of
the finally chosen ciphers by eSTREAM project, but it is weak under
soft fault analysis \cite{M2008,MH2008}.

   In this paper we consider another type of fault analysis of stream
cipher, which is to simplify the cipher system by injecting some
hard faults (that is, by setting the values of some positions
permanently 0). We call it hard fault analysis. Such attack was
presented by Eli~Biham and Adi~Shamir \cite{El1997}, used for
breaking block ciphers. We present the following results about hard
fault analysis of Trivium. In Case 1 with the probability not
smaller than 0.2396, the attacker can obtain 69 bits of
80--bits--key. In Case 2 with the probability not smaller than
0.2291, the attacker can obtain all of 80--bits--key. In Case 3 with
the probability not smaller than 0.2291, the attacker can partially
solve the key. In Case 4 with non--neglectable probability, the
attacker can obtain a simplified cipher, with smaller number of
state bits and slower non--linearization procedure. In Case 5 with
non--neglectable probability, the attacker can obtain another
simplified cipher. Besides, these 5 cases can be checked out by
observing the key--stream.

  The contents are organized as follows. Next subsection is an
explanation to soft fault analysis and hard fault analysis. In
section II we prepare for hard fault analysis of Trivium, including
description of Trivium, our assumptions, notations, and some facts.
In section III we present different features of fault injected
machine, in 7 different cases. In this section we show that, in each
of former 5 cases, either the key can be revealed, or the cipher can
be practically simplified. In section IV we present an algorithm to
identify the cases, by observing the key--stream. In this section we
identify the former 4 cases with the probability closed to 1, and
identify Case 5 with the probability no smaller than $4/5$. Section
V is the conclusion and future work expectation.
\subsection{Soft Fault Analysis and Hard Fault Analysis}
  Soft fault analysis is based on soft fault injection. At a random
moment of the encryption machine's driving procedure, the attacker
changes the values of some random positions of the state. By the
differential of the key--stream, the attacker can obtain several
additional low--degree--equations of the state.

  Hard fault analysis is based on hard fault injection. The attacker
makes the values of some random positions of the state permanently
0. That is, after hard fault injection, those injected bits can be
read out as 0, but can no longer be written in. According to
technical restriction, hard fault injection must be made before the
encryption machine's driving procedure.

 Three comparisons between hard fault analysis and soft fault analysis are as follows.

   Comparison 1:\hspace{0.05cm}  Hard fault analysis is more practical than soft fault
analysis. The main criticism against soft fault analysis was the
transient fault model that was claimed to be unrealistic
\cite{El1997}. Hard fault injection is a current technique for
micro--probing, and has already become real danger to cipher chip
\cite{Ro1997}. For example, DS5003 is a new product of Maxim. It is
a secure microprocessor chip by using coating technique, for
resisting hard fault injection.

  Comparison 2:\hspace{0.05cm}  Hard fault analysis is more expensive than soft fault
analysis. Soft fault injection is assumed to be made by simple fault
induction (special kind of light, magnetic disturbance, or other
brute methods). Hard fault injection needs expensive FIB and related
equipment.

  Comparison 3:\hspace{0.05cm}  After soft fault analysis, an encryption
machine can be returned back to the owner and be used again. On the
other hand, after hard fault analysis, an encryption machine is
destroyed, so that it seems meaningless to reveal the hidden key for
this machine. By this, it may be considered that hard fault analysis
is not as valuable as soft fault analysis. This may also be the
reason for that hard fault analysis has sparsely appeared in the
literature of stream cipher analysis.

    For Comparison 3, we argue
that hard fault analysis is useful in some application scenes. One
scene is that current key is used for decrypting the former
plain--texts before they are outdated. Another scene is that the
system has a weak key--renewal--algorithm, where current key can
help to predict future keys. The third scene is that several
machines share a common key, or have closely related keys.

\section{Preparation for Hard Fault Analysis of Trivium}
\subsection{Trivium Key--Stream Generation and Trivium State Initialization}
  The state of Trivium is 288 bits long, denoted as $(s_{1},\cdots,s_{288})$. The
state is renewed by 3 combined NFSRs (Non--linear Feedback Shift
Registers). The first NFSR is 93 bits long, denoted as
$(s_{1},\cdots, s_{93})$. The second NFSR is 84 bits long, denoted
as $(s_{94},\cdots, s_{177})$. The third NFSR is $111$ bits long,
denoted as $(s_{178},\cdots, s_{288})$. Current key--stream bit is a
linear function of current state. Table 1 is an equivalent algorithm
for the key--stream generation.
\begin{table}[!t]
\centering \caption{The key--stream generation algorithm}
\label{table_omega}
\begin{IEEEeqnarraybox}[\IEEEeqnarraystrutmode\IEEEeqnarraystrutsizeadd{2pt}{1pt}]{v/l/v}
\IEEEeqnarrayrulerow\\
&\mbox{Input: the initial state $(s_{1},\cdots,s_{288})$,\hspace{2.5cm}}&\\
&\mbox{\hspace{0.75cm}the number
of output bits $N\leq 2^{64}$}&\\
&\mbox{Output: key-stream $(z_{0},z_{1},z_{2},\cdots,z_{N})$}&\\
\IEEEeqnarrayrulerow\\
\IEEEeqnarrayseprow[3pt]\\
&1:for~ i=0~  to~  N-1~ do&\IEEEeqnarraystrutsize{0pt}{0pt}\\
\IEEEeqnarrayseprow[3pt]\\
&2:z_{i}\leftarrow
   s_{66}+s_{93}+s_{162}+s_{177}+s_{243}+s_{288} &\IEEEeqnarraystrutsize{0pt}{0pt}\\
\IEEEeqnarrayseprow[3pt]\\
&3:t_{1}\leftarrow
       s_{66}+s_{91}s_{92}+s_{93}+s_{171}&\IEEEeqnarraystrutsize{0pt}{0pt}\\
\IEEEeqnarrayseprow[3pt]\\
&4:t_{2}\leftarrow
          s_{162}+s_{175}s_{176}+s_{177}+s_{264}&\IEEEeqnarraystrutsize{0pt}{0pt}\\
\IEEEeqnarrayseprow[3pt]\\
&5:t_{3}\leftarrow
      s_{243}+s_{286}s_{287}+s_{288}+s_{69}&\IEEEeqnarraystrutsize{0pt}{0pt}\\
\IEEEeqnarrayseprow[3pt]\\
&6:(s_{1},\cdots,s_{93})\leftarrow
      (t_{3},s_{1},\cdots,s_{92})&\IEEEeqnarraystrutsize{0pt}{0pt}\\
\IEEEeqnarrayseprow[3pt]\\
&7:(s_{94},\cdots,s_{177})\leftarrow
          (t_{1},s_{94},\cdots,s_{176})&\IEEEeqnarraystrutsize{0pt}{0pt}\\
\IEEEeqnarrayseprow[3pt]\\
&8:(s_{178},\cdots,s_{288})\leftarrow
    (t_{2},s_{178},\cdots,s_{287})&\IEEEeqnarraystrutsize{0pt}{0pt}\\
\IEEEeqnarrayseprow[3pt]\\
&9:end ~for&\IEEEeqnarraystrutsize{0pt}{0pt}\\
\IEEEeqnarrayrulerow
\end{IEEEeqnarraybox}
\end{table}

    The key is 80 bits long, denoted as $(k_{1},\cdots, k_{80})$, and is secret. IV
(Initial Vector) is 80 bits long, denoted as $(IV_{1},\cdots,
IV_{80})$, and is public. In other words, if anyone obtains an
encryption machine, he can arbitrarily set the value of IV. Table 2
is an equivalent algorithm for the initial state generation.
\begin{table}[!t]
\centering \caption{The initial state generation algorithm}
\label{table_omega}
\begin{IEEEeqnarraybox}[\IEEEeqnarraystrutmode\IEEEeqnarraystrutsizeadd{2pt}{1pt}]{v/l/v}
\IEEEeqnarrayrulerow\\
&\mbox{Input: the  state }&\\
\IEEEeqnarrayseprow[3pt]\\
&(s_{1},\cdots,s_{93})\leftarrow
      (k_{1},\cdots,k_{80},0,\cdots,0)&\IEEEeqnarraystrutsize{0pt}{0pt}\\
\IEEEeqnarrayseprow[3pt]\\
&(s_{94},\cdots,s_{177})\leftarrow
          (IV_{1},\cdots,IV_{80},0,\cdots,0)&\IEEEeqnarraystrutsize{0pt}{0pt}\\
\IEEEeqnarrayseprow[3pt]\\
&(s_{178},\cdots,s_{288})\leftarrow
    (0,\cdots,0,1,1,1)&\IEEEeqnarraystrutsize{0pt}{0pt}\\
&\mbox{Output: the initial state $(s_{1},\cdots,s_{288})$}&\\
\IEEEeqnarrayrulerow\\
\IEEEeqnarrayseprow[3pt]\\
&1:for~ $i=1$~ to ~$1152$~ do&\IEEEeqnarraystrutsize{0pt}{0pt}\\
\IEEEeqnarrayseprow[3pt]\\
&2:t_{1}\leftarrow
       s_{66}+s_{91}s_{92}+s_{93}+s_{171} &\IEEEeqnarraystrutsize{0pt}{0pt}\\
\IEEEeqnarrayseprow[3pt]\\
&3:t_{2}\leftarrow
          s_{162}+s_{175}s_{176}+s_{177}+s_{264}&\IEEEeqnarraystrutsize{0pt}{0pt}\\
\IEEEeqnarrayseprow[3pt]\\
&4:t_{3}\leftarrow
      s_{243}+s_{286}s_{287}+s_{288}+s_{69}&\IEEEeqnarraystrutsize{0pt}{0pt}\\
\IEEEeqnarrayseprow[3pt]\\
&5:(s_{1},\cdots,s_{93})\leftarrow
      (t_{3},s_{1},\cdots,s_{92})&\IEEEeqnarraystrutsize{0pt}{0pt}\\
\IEEEeqnarrayseprow[3pt]\\
&6:(s_{94},\cdots,s_{177})\leftarrow
          (t_{1},s_{94},\cdots,s_{176})&\IEEEeqnarraystrutsize{0pt}{0pt}\\
\IEEEeqnarrayseprow[3pt]\\
&7:(s_{178},\cdots,s_{288})\leftarrow
    (t_{2},s_{178},\cdots,s_{287})\hspace{1.45cm}&\IEEEeqnarraystrutsize{0pt}{0pt}\\
\IEEEeqnarrayseprow[3pt]\\
&8:end~ for&\IEEEeqnarraystrutsize{0pt}{0pt}\\
\IEEEeqnarrayrulerow
\end{IEEEeqnarraybox}
\end{table}

   Table 1 and Table 2 show that, for key--stream generation and initial
state generation, the state renewal is the same. In detail, let
$s_{(t,j)}$denote the state bit at time $t$ and position $j$, then
Table 3 presents a clearer description for the state renewal.
\begin{table}[!t]
\centering \caption{The state renewal}
\label{table_omega}
\begin{IEEEeqnarraybox}[\IEEEeqnarraystrutmode\IEEEeqnarraystrutsizeadd{2pt}{1pt}]{v/l/v}
\IEEEeqnarrayrulerow\\
\IEEEeqnarrayseprow[3pt]\\
&\hspace{0.3cm}(s_{(t+1,1)},s_{(t+1,2)},\cdots,s_{(t+1,93)})&\IEEEeqnarraystrutsize{0pt}{0pt}\\
\IEEEeqnarrayseprow[3pt]\\
&=(s_{(t,243)}+s_{(t,286)}s_{(t,287)}+s_{(t,288)}+s_{(t,69)},s_{(t,1)},&\\
\IEEEeqnarrayseprow[3pt]\\
&\hspace{0.3cm}s_{(t,1)},\cdots,s_{(t,92)})&
\IEEEeqnarraystrutsize{0pt}{0pt}\\
      \IEEEeqnarrayrulerow\\
\IEEEeqnarrayseprow[3pt]\\
&\hspace{0.3cm}(s_{(t+1,94)},s_{(t+1,95)},\cdots,s_{(t+1,177)})&
\IEEEeqnarraystrutsize{0pt}{0pt}\\
\IEEEeqnarrayseprow[3pt]\\
&=(s_{(t,66)}+s_{(t,91)}s_{(t,92)}+s_{(t,93)}+s_{(t,171)},&\\
\IEEEeqnarrayseprow[3pt]\\
&\hspace{0.3cm}s_{(t,94)},\cdots,s_{(t,176)})&
\IEEEeqnarraystrutsize{0pt}{0pt}\\
\IEEEeqnarrayrulerow\\
\IEEEeqnarrayseprow[3pt]\\
&\hspace{0.3cm}(s_{(t+1,178)},s_{(t+1,179)},\cdots,s_{(t+1,288)})&\IEEEeqnarraystrutsize{0pt}{0pt}\\
\IEEEeqnarrayseprow[3pt]\\
&=(s_{(t,162)}+s_{(t,175)}s_{(t,176)}+s_{(t,177)}+s_{(t,264)},&\\
\IEEEeqnarrayseprow[3pt]\\
&\hspace{0.3cm}s_{(t,178)},\cdots,s_{(t,287)})&
\IEEEeqnarraystrutsize{0pt}{0pt}\\
\IEEEeqnarrayrulerow
\end{IEEEeqnarraybox}
\end{table}
\newtheorem{lemma}{Lemma}
 \begin{lemma}
  \cite{C2005,CD}
Let $(s_{1},\cdots,s_{288})$ denote the initial state
 (that is, the state at the time just before generating $z_{0}$).
 Take $\{z_{0},z_{1},z_{2},\cdots\}$ as functions of $(s_{1},\cdots,s_{288})$. Then
\begin{enumerate}
  \item $\{z_{0},z_{1},\cdots,z_{65}\}$ are 66 linear functions.
  \item $\{z_{66},z_{67},\cdots,z_{147}\}$ are 82 quadratic functions.
  \item $\{z_{148},z_{149},\cdots,z_{213}\}$ are 66 cubic functions.
  \item Each of $\{z_{214},z_{215},\cdots,\}$ is at least a quartic function.
\end{enumerate}
\end{lemma}

   Lemma 1 shows such a weakness of Trivium that its non--linearization procedure
 is over slow. By knowing the key--stream, a large number of low--degree--equations
 will be obtained.
\subsection{Assumptions, Notations and Some Facts}
  Suppose that the attacker obtains an encryption machine (or an
encryption card, etc), equipped with Trivium. He wants to obtain the
hidden key $(k_{1},\cdots, k_{80})$. He makes hard fault injection.
The hard fault bits are from random one of 3 NFSRs, and at random
positions in this NFSR. At injecting moment, he can not control the
positions of hard fault bits. After injection, he does not know the
positions of hard fault bits. Then he set$(IV_{1},\cdots, IV_{80})=
(0,\cdots, 0)$. That is, for initial state generation procedure, the
input state is

$(s_{1},\cdots, s_{93}) \leftarrow
(k_{1},\cdots,k_{80},0,\cdots,0),$

$(s_{94},\cdots, s_{177})\leftarrow (0,\cdots,0),$

$(s_{178},\cdots, s_{288})\leftarrow (0,\cdots,0,1,1,1).$\\
  Then he starts up the machine (initial state generation and key--stream
generation), and checks the output key--stream from this
fault--injected machine.

  It is easy to see that our assumptions are quite trivial.

   $P_{L}$ denotes the lowest position of injected faults. $P_{H}$ denotes the highest
position of injected faults. According to our assumptions, $P_{H}$
and $P_{L}$ fall into the same index set $\{1,\cdots, 93\}$, or $
\{94,\cdots, 177\}$, or $\{178,\cdots, 288\}$. $P_{L}$ is of the
following 7 cases.

  Case 1:\hspace{0.1cm}$94\leq P_{L}\leq 162$.

  Case 2:\hspace{0.1cm}$178\leq P_{L}\leq 243$.

  Case 3:\hspace{0.1cm}$1\leq P_{L}\leq 66$.

  Case 4:\hspace{0.1cm}$163\leq P_{L}\leq 171$.

  Case 5:\hspace{0.1cm}$172\leq P_{L}\leq 176$.

  Case 6:\hspace{0.1cm}$ P_{L}=177$.

  Case 7:\hspace{0.1cm}other values of $ P_{L}$, that is,

   \hspace{1cm}$67\leq P_{L}\leq 93$ or $244\leq P_{L}\leq 288 $.

   It is clear that the probability of Case 1 is never smaller than 69/288=0.2396,
that the probability of Case 2 is never smaller than 66/288=0.2291,
and that the probability of Case 3 is never smaller than
66/288=0.2291. Probabilities of Case 4 and Case 5 are not clear,
because we do not set detailed injection model. We can only say that
these 2 probabilities are non--neglectable. The probability of Case
6 is never larger than 1/288=0.0035, and generally is far smaller
than 0.0035.

  We call the input state the state at time 0, and sequentially rank
the state at time $1,2,\cdots$. By this ranking, the initial state
(that is, the state at the time just before generating $z_{0}$) is
the state at time 1152. $(s_{(t, 1)}, s_{(t, 2)},\cdots, s_{(t,
288)})$ denotes the state at time $t$. So that, for each $m\geq0$,
the key--stream bit $z_{m}$ has such a ~representation~

 $z_{m}=s_{(m+1152,66)}+s_{(m+1152,93)}+s_{(m+1152,162)}$

 $ {\ \ \ \ }+s_{(m+1152,177)}+s_{(m+1152,243)}+s_{(m+1152,288)}.$\vspace{0.2cm}

 $\ast$ denotes an arbitrary bit--value.

   Some simple facts about hard fault injection are as follows.

  Suppose $j$ is a position of hard fault injected bit, where $1\leq j\leq 93$.
Then $s_{(t,j+m)}=0$ for each $(t, m)$ such that $t\geq 0$ and
$0\leq m\leq min\{93-j, t\}$.

  Suppose $j$ is a position of hard fault injected bit, where $94\leq j\leq 177$.
 Then $s_{(t,j+m)}=0$ for each $(t, m)$ such that $t\geq 0$ and $0\leq m\leq min\{177-j, t\}$.

  Suppose $j$ is a position of hard fault injected bit, where$178\leq j\leq 288$.
Then $s_{(t,j+m)}=0$ for each $(t, m)$ such that $t\geq 0$ and
$0\leq m\leq min\{288-j, t\}$.

\section{Features of Fault Injected Machine in 7 Cases}
\subsection{Features of Fault Injected Machine in
  Case 1:~$94\leq P_{L}\leq 162$}
 \begin{lemma}
  The state at time 27 is the follow.
  \begin{enumerate}
   \item  \hspace{0.4cm}$(s_{(27,1)},\cdots,s_{(27,93)})$

  \hspace{0.1cm}$=(k_{43},\cdots,k_{66},k_{67}+1,k_{68}+1,k_{69},k_{1},\cdots,k_{66}).$
   \item   \hspace{0.1cm}$(s_{(27,94)},\cdots,s_{(27,161)})=(\ast,\cdots,\ast),$
and

     \hspace{0.1cm}$(s_{(27,162)},\cdots,s_{(27,177)})=(0,\cdots,0)$.
   \item \hspace{0.1cm}$(s_{(27,178)},\cdots,s_{(27,288)})=(0,\cdots,0)$.
    \end{enumerate}
     \end{lemma}\vspace{0.2cm}

\begin{lemma}{\ }
\begin{enumerate}
  \item  For each $t$ such that $t\geq 27$,

$(s_{(t+1,1)},\cdots,s_{(t+1,93)})=(s_{(t,69)},s_{(t,1)},\cdots,s_{(t,92)})$.\\
 So that $ \{(s_{(t,1)},\cdots,s_{(t,93)}),t\geq27\}$has a period 69.
  \item  For each $t$ such that $t\geq 27$,

$(s_{(t,70)},\cdots,s_{(t,93)})=(s_{(t,1)},\cdots,s_{(t,24)})$.
  \item For each $t$ such that $t\geq 27$,

$(s_{(t,162)},\cdots,s_{(t,288)})=(0,\cdots,0)$.
\end{enumerate}
\end{lemma}\vspace{0.2cm}

  Lemma 2 and Lemma 3 are clear by gradually renewing the state (see
  Table 3), and by considering the state at time 0:

  $(s_{(0,1)},\cdots,s_{(0,93)})=(k_{1},\cdots,k_{80},0,\cdots,0).$

  $(s_{(0,94)},\cdots,s_{(0,177)})=(0,\cdots,0)$.

  $(s_{(0,178)},\cdots,s_{(0,288)})=(0,\cdots,0,1,1,1)$.\vspace{0.2cm}

\newtheorem{proposition}{Proposition}
\begin{proposition}
  Suppose $94\leq P_{L}\leq162$. Then the key--stream $(z_{0}z_{1}z_{2}\cdots)$ has a period $69$,
  where

  \hspace{0.5cm}$(z_{0},z_{1},z_{2},\cdots,z_{68})$

  $=(k_{18},k_{17},\cdots,k_{1},k_{69},k_{68}+1,k_{67}+1,k_{66},k_{65},\cdots,k_{19}).$
\end{proposition}
\renewcommand{\IEEEQED}{\IEEEQEDopen}
\begin{IEEEproof}
  By Lemma 2 and Lemma 3,
  $z_{0}=s_{(1152,66)}$,
$z_{1}=s_{(1153,66)},z_{2}=s_{(1154,66)}\cdots$. So that the
key--stream $(z_{0}z_{1}z_{2}\cdots)$ has a period 69. Again
$z_{0}=s_{(1152,66)}=s_{(27,45)}=k_{18}$.
  Proposition 1 is proved.
\end{IEEEproof}
\subsection{Features of Fault Injected Machine in
  Case 2:$178\leq P_{L}\leq 243$}
\begin{lemma}
  The state at time 27 is the follow.
\begin{enumerate}
  \item \hspace{0.4cm}$(s_{(27,1)},\cdots,s_{(27,93)})$

    $=(k_{43},\cdots,k_{66},k_{67}+1,k_{68}+1,k_{69},k_{1},\cdots,k_{66}).$
  \item \hspace{0.4cm}$(s_{(27,94)},\cdots,s_{(27,177)})$

   $=(k_{40}+k_{65}k_{66}+k_{67},k_{41}+k_{66}k_{67}+k_{68},\cdots,$

 \hspace{0.4cm}$k_{53}+k_{78}k_{79}+k_{80},k_{54}+k_{79}k_{80},$

  \hspace{0.4cm}$k_{55},k_{56},\cdots,k_{66},0,\cdots,0).$
  \item $(s_{(27,178)},\cdots,s_{(27,242)})=(\ast,\cdots,\ast).$
  \item $(s_{(27,243)},\cdots,s_{(27,288)})=(0,\cdots,0)$.
\end{enumerate}
\end{lemma}
\begin{IEEEproof}
  We induce the state at time 27 by gradually renewing the state.

  The state at time 1:

  $(s_{(1,1)},\cdots,s_{(1,93)})=(k_{69},k_{1},\cdots,k_{80},0,\cdots,0)$,

 $(s_{(1,94)},\cdots,s_{(1,177)})=(k_{66},0,\cdots,0)$,

 $(s_{(1,178)},\cdots,s_{(1,288)})=(0,\cdots,0,1,1)$.

 The state at time 2:

 $(s_{(2,1)},\cdots,s_{(2,93)})=(k_{68}+1,k_{69},k_{1},\cdots,k_{80},0,\cdots,0)$,

 $(s_{(2,94)},\cdots,s_{(2,177)})=(k_{65},k_{66},0,\cdots,0)$,

 $(s_{(2,178)},\cdots,s_{(2,288)})=(0,\cdots,0,1)$.

 The state at time 3:

 \hspace{0.3cm}$(s_{(3,1)},\cdots,s_{(3,93)})$

$=(k_{67}+1,k_{68}+1,k_{69},k_{1},\cdots,k_{80},0,\cdots,0)$,

 $(s_{(3,94)},\cdots,s_{(3,177)})=(k_{64},k_{65},k_{66},0,\cdots,0)$,

 $(s_{(3,178)},\cdots,s_{(3,288)})=(0,\cdots,0)$.

 The state at time 12:

 \hspace{0.3cm}$(s_{(12,1)},\cdots,s_{(12,93)})$

 $=(k_{58},\cdots,k_{66},k_{67}+1,k_{68}+1,k_{69},k_{1},\cdots,k_{80},0)$,

 $(s_{(12,94)},\cdots,s_{(12,177)})=(k_{55},\cdots,k_{66},0,\cdots,0)$,

 $(s_{(12,178)},\cdots,s_{(12,242)})=(\ast,\cdots,\ast)$,

$(s_{(12,243)},\cdots,s_{(12,288)})=(0,\cdots,0)$.

The state at time 13:

 \hspace{0.3cm}$(s_{(13,1)},\cdots,s_{(13,93)})$

 $=(k_{57},\cdots,k_{66},k_{67}+1,k_{68}+1,k_{69},k_{1},\cdots,k_{80})$,

 \hspace{0.3cm}$(s_{(13,94)},\cdots,s_{(13,177)})$

 $=(k_{54}+k_{79}k_{80},k_{55},\cdots,k_{66},0,\cdots,0)$,

 $(s_{(13,178)},\cdots,s_{(13,242)})=(\ast,\cdots,\ast)$,

$(s_{(13,243)},\cdots,s_{(13,288)})=(0,\cdots,0)$.

The state at time 14:

 \hspace{0.3cm}$(s_{(14,1)},\cdots,s_{(14,93)})$

$=(k_{56},\cdots,k_{66},k_{67}+1,k_{68}+1,k_{69},k_{1},\cdots,k_{79})$,

\vspace{0.1cm}
 \hspace{0.3cm}$(s_{(14,94)},\cdots,s_{(14,177)})$

$=(k_{53}+k_{78}k_{79}+k_{80},k_{54}+k_{79}k_{80},k_{55},\cdots,$

  \hspace{0.4cm}$k_{66},0,\cdots,0)$,

 $(s_{(14,178)},\cdots,s_{(14,242)})=(\ast,\cdots,\ast)$,

$(s_{(14,243)},\cdots,s_{(14,288)})=(0,\cdots,0)$.

The state at time 27:

 \hspace{0.3cm}$(s_{(27,1)},\cdots,s_{(27,93)})$

$=(k_{43},\cdots,k_{66},k_{67}+1,k_{68}+1,k_{69},k_{1},\cdots,k_{66})$,

 \hspace{0.3cm}$(s_{(27,94)},\cdots,s_{(27,177)})$

 $=(k_{40}+k_{65}k_{66}+k_{67},\cdots,k_{53}+k_{78}k_{79}+k_{80},k_{54}+k_{79}k_{80},$

  \hspace{0.4cm}$k_{55},\cdots,k_{66},0,\cdots,0)$,

 $(s_{27,178)},\cdots,s_{(27,242)})=(\ast,\cdots,\ast)$,

$(s_{(27,243)},\cdots,s_{(27,288)})=(0,\cdots,0)$.

Lemma 4 is proved.
\end{IEEEproof} \vspace{0.2cm}

Notice that 1) and 2) of Lemma 3 are still true for  Case2: $178\leq
P_{L}\leq 243$. Now we present a definition. For each $t$ such that
$t\geq 27$, define
$a_{t+1}=s_{(t,66)}+s_{(t,91)}s_{(t,92)}+s_{(t,93)}$. For each $t$
such that $0\leq t<27$, define $a_{t+1}=a_{t+70}$.\vspace{0.2cm}

\begin{lemma}{\ }
\begin{enumerate}
  \item For each $t$ such that $t\geq 27$,

\hspace{1.4cm}$(s_{(t+1,94)},\cdots,s_{(t+1,177)})$

\hspace{1cm}$=(s_{(t,177)}+a_{t+1},s_{(t,94)},\cdots,s_{(t,176)}).$

  \item $\{a_{t+1},t\geq 27\}$ has a period 69, where

 $ (a_{28},\cdots,a_{96})=(k_{39}+k_{64}k_{65}+k_{66},k_{38}+k_{63}k_{64}+k_{65},\cdots,$
 $k_{1}+k_{26}k_{27}+k_{28},k_{69}+k_{25}k_{26}+k_{27},k_{68}+1+k_{24}k_{25}+k_{26},$
 $k_{67}+1+k_{23}k_{24}+k_{25},k_{66}+k_{22}k_{23}+k_{24},k_{65}+k_{21}k_{22}+k_{23},\cdots,$
 $k_{45}+k_{1}k_{2}+k_{3},k_{44}+k_{69}k_{1}+k_{2},k_{43}+(k_{68}+1)k_{69}+k_{1},$
 $k_{42}+(k_{67}+1)(k_{68}+1)+k_{69},k_{41}+k_{66}(k_{67}+1)+k_{68}+1,$
 $k_{40}+k_{65}k_{66}+k_{67}+1).$
  \item $\{a_{t+1},t\geq 27\}$ has a period 69.
\end{enumerate}
\end{lemma}
\begin{IEEEproof}
 1) is clear from Trivium state renewal.
For each $t$ such that $t\geq 27$, each $j$ such that $1\leq j\leq
69,s_{(t,j)}=s_{(27,j-t+27(mod 69))}$ . So that
\begin{displaymath}
\left.
\begin{array}{ll}
a_{t+1}\hspace{-0.3cm}&=s_{(t,66)}+s_{(t,91)}s_{(t,92)}+s_{(t,93)}\\
       &=s_{(t,66)}+s_{(t,22)}s_{(t,23)}+s_{(t,24)}\\
       &=s_{(27,24-t(mod69))}+s_{(27,49-t(mod69))}s_{(27,50-t(mod69))}\\
       &\hspace{0.3cm}+s_{(27,51-t(mod69))}.
  \end{array}\right.
  \end{displaymath}

  So that 2) is true, and 3) is immediate from 2). Lemma 5 is proved.
  \end{IEEEproof}
\vspace{0.2cm}

\begin{lemma}
 Take the following changes for the state at time 27.
$(s_{(27,172)},\cdots,s_{(27,177)})$ are changed as

  \hspace{1cm}$(s_{(27,172)},\cdots,s_{(27,177)})$

\hspace{0.6cm}$=(s_{(27,94)}+a_{27},s_{(27,95)}+a_{26},\cdots,s_{(27,99)}+a_{22}),$\\
and other positions of the state at time 27 are kept unchanged. Then
\begin{enumerate}
  \item  For each $t$ such that $t\geq33$,
  $(s_{(t,1)},\cdots,s_{(t,177)})$ and
  $(s_{(t,243)},\cdots,s_{(t,288)})$ are kept unchanged.

  \item The key--stream $(z_{0}z_{1}z_{2}\cdots)$ are kept unchanged.
\end{enumerate}
\end{lemma}
\begin{IEEEproof}
 \emph{Proof:}\hspace{0.1cm}Notice that we are in Case 2: $178\leq P_{L}\leq243$, and that the
 state bits shift rightwards. So that Lemma 6 is clear.
 \end{IEEEproof}
\vspace{0.2cm}

\begin{lemma}
 Take the state at time $27$ as the changed value as described in Lemma 6.
 Then For each $t$ such that $t\geq 27$, each $j$ such that
  $94\leq j \leq177, s_{(t+78, j)}=s_{(t, j)}+a_{t+172-j}$.
  \end{lemma}
\begin{IEEEproof}
\begin{enumerate}
  \item If $94\leq j \leq171$ and $t\geq 27$, then $t+172-j\geq 28$, so
  that
\begin{displaymath}
\left.
\begin{array}{ll}
s_{(t+78,j)}\hspace{-0.3cm}&=s_{(t+172-j,94)}\\
       &=s_{(t+171-j,171)}+a_{t+172-j}\\
       &=s_{(t,j)}+a_{t+172-j}.
  \end{array}\right.
  \end{displaymath}

  \item If $172\leq j \leq177$ and $t\geq 33$, then $156\leq j-6\leq171$ and $t-6\geq
 27$. By 1),
\begin{displaymath}
\left.
\begin{array}{ll}
s_{(t+78,j)}\hspace{-0.3cm}&=s_{(t-6+78,j-6)}\\
       &=s_{(t-6,j-6)}+a_{t-6+172-(j-6)}\\
       &=s_{(t,j)}+a_{t+172-j}.
  \end{array}\right.
  \end{displaymath}

  \item If $172\leq j \leq177$ and $t=27$, then $94\leq j-78\leq99$,
  so that
$s_{(27+78,j)}=s_{(27,j-78)}$. By the assumptions of Lemma 6,
\begin{displaymath}
\left.
\begin{array}{ll}
s_{(27+78,j)}\hspace{-0.3cm}&=s_{(27,j-78)}\\
       &=s_{(27,j)}+a_{27+172-j}.
  \end{array}\right.
  \end{displaymath}

  \item If $172\leq j \leq177,28\leq t\leq32$, and $j-(t-27)\leq171$,
then $167\leq j-(t-27)\leq 171$. By 1),
\begin{displaymath}
\left.
\begin{array}{ll}
s_{(t+78,j)}\hspace{-0.3cm}&=s_{(27+78,j-(t-27))}\\
       &=s_{(27,j-(t-27))}+a_{27+172-(j-(t-27))}\\
       &=s_{(t,j)}+a_{t+172-j}.
  \end{array}\right.
  \end{displaymath}

  \item If $172\leq j \leq177,28\leq t\leq32$, and $j-(t-27)\geq172$,
then $172\leq j-(t-27)\leq 176$. By 3),
\begin{displaymath}
\left.
\begin{array}{ll}
s_{(t+78,j)}\hspace{-0.3cm}&=s_{(27+78,j-(t-27))}\\
       &=s_{(27,j-(t-27))}+a_{27+172-(j-(t-27))}\\
       &=s_{(t,j)}+a_{t+172-j}.
  \end{array}\right.
  \end{displaymath}
\end{enumerate}
  Lemma 7 is proved.
  \end{IEEEproof}
\vspace{0.2cm}

\begin{lemma}
  Take the state at time 27 as the changed value as described in Lemma 6. Then
\begin{enumerate}
  \item For each $t$ such that $t\geq27$, each $j$ such that

$94\leq j\leq177$,
$$s_{(t+1794,j)}=s_{(t,j)}+\sum^{22}_{m=0}a_{t+34-j+3m}.$$

  \item $\{(s_{(t, 1)},\cdots, s_{(t, 177)}),t\geq27\}$ has a period
3358.
\end{enumerate}
\end{lemma}
\begin{IEEEproof}
According to Lemma 5, Lemma 6, Lemma 7 and the fact that
$1794=78\times23=69\times26$,
 \begin{displaymath}
 \left.
 \begin{array}{ll}
  s_{(t+1794,j)}\hspace{-0.3cm}&=s_{(t+78\times 23,j)}\vspace{0.2cm}\\
       &=s_{(t,j)}+\sum^{22}_{n=0}a_{t+172-j+78\times n(mod69)}\vspace{0.2cm}\\
       &=s_{(t,j)}+\sum^{22}_{m=0}a_{t+34-j+3m},
  \end{array}\right.
  \end{displaymath}
 so that 1) is true. According to 1), for each $t$ such that
  $t\geq27$, each $j$such that $94\leq j\leq 177$,
 \begin{displaymath}
 \left.
 \begin{array}{ll}
  s_{(t+3588,j)}\hspace{-0.3cm}&=s_{(t+1794+1794,j)}\vspace{0.2cm}\\
       &=s_{(t,j)}+\sum^{22}_{m=0}a_{t+34-j+3m}\vspace{0.2cm}\\
       &\hspace{0.2cm}+\sum^{22}_{m=0}a_{t+1794+34-j+3m}\\
       &=s_{(t,j)}.
  \end{array}\right.
  \end{displaymath}
  This implies that $\{(s_{(t, 94)},\cdots,s_{(t, 177)}), t\geq27\}$ has
  a period 3358. Again by the fact that  $\{(s_{(t,1)},\cdots,s_{(t,93)}),t\geq27\}$
  has a period 69, 2) is true. Lemma 8 is proved.
  \end{IEEEproof}
\vspace{0.2cm}

  \begin{proposition}
  Suppose $178\leq P_{L}\leq243$. Then
\begin{enumerate}
  \item  The key--stream $(z_{0}z_{1}z_{2}\cdots)$ has a period 3358.

  \item $\{z_{0},z_{1},z_{2},\cdots,z_{3357}\}$ are linear functions of
216
  variables

  $(s_{(27,25)},\cdots,s_{(27,93)},s_{(27,100)},\cdots,s_{(27,177)},a_{28},\cdots,a_{96}),$
 and these functions are known.

  \item By knowing the values of $\{z_{0},z_{1},z_{2},\cdots,z_{3357}\}$,
 the attacker obtains 3358 linear equations of 216 variables

$(s_{(27,25)},\cdots,s_{(27,93)},s_{(27,100)},\cdots,s_{(27,177)},a_{28},\cdots,a_{96}).$
 The rank of these linear equations is 210, so that there are $2^6=64$ possible solutions.
\end{enumerate}
 \end{proposition}
 \begin{IEEEproof}
 1) is clear from Lemma 8. Notice that for each $t$ such that $t\geq 27$,

 $(s_{(t+1,1)},\cdots,s_{(t+1,93)})=(s_{(t,69)},s_{(t,1)},\cdots,s_{(t,92)}),$

 \hspace{1.4cm}$(s_{(t+1,94)},\cdots,s_{(t+1,177)})$

\hspace{1cm}$=(s_{(t,171)}+a_{t+1},s_{(t,94)},\cdots,s_{(t,176)}).$

So that, for each $t$ such that
$t\geq27$,$(s_{(t,1)},\cdots,s_{(t,177)})$ can be induced from

$(s_{(27,25)},\cdots,s_{(27,93)},s_{(27,100)},\cdots,s_{(27,177)},a_{28},\cdots,a_{96})$
by linear recursion which is already known. So that 2) is true.

3) is our checking result. Proposition 2 is proved.
 \end{IEEEproof}
\vspace{0.2cm}

 Notice that the true value of

 $(s_{(27,25)},\cdots,s_{(27,93)},s_{(27,100)},\cdots,s_{(27,177)},a_{28},\cdots,a_{96})$
 satisfies

\hspace{0.8cm} $(s_{(27,25)},\cdots,s_{(27,93)})$

 \hspace{0.4cm}$=(k_{67}+1,k_{68}+1,k_{69},k_{1},\cdots,k_{66}),$

 \hspace{-0.2cm}and

$(s_{(27,100)},\cdots,s_{(27,177)})=(k_{46}+k_{71}k_{72}+k_{73},\cdots,$

 $k_{53}+k_{78}k_{79}+k_{80},k_{54}+k_{79}k_{80},k_{55},\cdots,k_{66},0,\cdots,0,$

$k_{40}+k_{65}k_{66}+k_{67}+a_{27},\cdots,k_{45}+k_{70}k_{71}+k_{72}+a_{22}).$

These relations present another group of equations of 216 variables

\hspace{-0.15cm}$(s_{(27,25)},\cdots,s_{(27,93)},s_{(27,100)},\cdots,s_{(27,177)},a_{28},\cdots,a_{96}),$
 described as the follow.

 \hspace{1cm}$(s_{(27,109)},\cdots,s_{(27,171)})$

 \hspace{0.6cm}$=(s_{(27,82)},\cdots,s_{(27,93)},0,\cdots,0),$

 $a_{28}=s_{(27,66)}+s_{(27,91)}s_{(27,92)}+s_{(27,93)},$

 $a_{29}=s_{(27,65)}+s_{(27,90)}s_{(27,91)}+s_{(27,92)},$

 $\cdots$

 $a_{69}=s_{(27,25)}+s_{(27,50)}s_{(27,51)}+s_{(27,52)},$

 $a_{70}=s_{(27,93)}+s_{(27,49)}s_{(27,50)}+s_{(27,51)},$

 $a_{71}=s_{(27,92)}+s_{(27,48)}s_{(27,49)}+s_{(27,50)},$

 $\cdots$

 $a_{94}=s_{(27,69)}+s_{(27,25)}s_{(27,26)}+s_{(27,27)},$

 $a_{95}=s_{(27,68)}+s_{(27,93)}s_{(27,25)}+s_{(27,26)},$

 $a_{96}=s_{(27,67)}+s_{(27,92)}s_{(27,93)}+s_{(27,25)}.$
\vspace{0.2cm}

 All these equations are enough to determine the true value of
$(s_{(27,25)},\cdots,s_{(27,93)},s_{(27,100)},\cdots,s_{(27,177)},a_{28},\cdots,a_{96}),$
so that enough to determine the value of $(k_{1},\cdots,k_{69})$.
Besides, all these equations can determine the value of

\hspace{0.4cm}$(k_{68}k_{69}+k_{70},k_{69}k_{70}+k_{71},\cdots,k_{78}k_{79}+k_{80}),$\\
so that determine the value of $(k_{70},\cdots,k_{80})$.
 \subsection{Features of Fault Injected Machine in Case 3:
  $1\leq P_{L}\leq 66$}
\begin{lemma}{\ }
\begin{enumerate}
  \item For each $t$ such that $t\geq92$,

 \hspace{1.4cm}$(s_{(t,66)},\cdots,s_{(t,93)})=(0,\cdots,0).$

  \item For each $t$ such that $t\geq98$,

 \hspace{0.7cm}$(s_{(t,172)},\cdots,s_{(t,177)})=(s_{(t,94)},\cdots,s_{(t,99)}).$

  \item $\{(s_{(t,94)},\cdots,s_{(t,177)}),t\geq98$ has a period
78.
\end{enumerate}
\end{lemma}
 \begin{IEEEproof}
 1) is clear in Case 3. 2) and 3) are immediate from 1).
 \end{IEEEproof}
\vspace{0.2cm}

  Now we present a definition. For each $t$ such that $t\geq98$, define

 \hspace{1cm}$b_{t+1}=s_{(t,162)}+s_{(t,175)}s_{(t,176)}+s_{(t,177)}.$

 For each $t$
such that $0\leq t<98$, define $b_{t+1}=b_{t+79}$. \vspace{0.2cm}

\begin{lemma}{\ }
\begin{enumerate}
  \item For each $t$ such that $t\geq98$,

  \hspace{0.7cm}$(s_{(t+1,178)},\cdots,s_{(t+1,288)})$

 \hspace{0.3cm}$=(s_{(t,264)}+b_{t+1},(s_{(t,178)}\cdots,s_{(t,287)}).$

  \item $\{b_{t+1},t\geq 0\}$has a period 78.
\end{enumerate}
\end{lemma}
 \begin{IEEEproof}
 Lemma 10 is just similar to Lemma 5.
  \end{IEEEproof}
\vspace{0.2cm}

\begin{lemma}
Take the following changes for the state at time 98.
$(s_{(98,265)},\cdots,s_{(98,288)})$ are changed as

\hspace{0.4cm}$(s_{(98,265)},\cdots,s_{(98,288)})$

 $=(s_{(98,178)}+b_{98},s_{(98,179)}+b_{97},\cdots,s_{(98,201)}+b_{75}),$\\
and other positions of the state at time 98 are kept unchanged. Then
\begin{enumerate}
  \item For each $t$ such that $t\geq122$,
$(s_{(t,66)},\cdots,s_{(t,288)})$ are kept unchanged.

  \item The key--stream $(z_{0}z_{1}z_{2}\cdots)$  are kept unchanged.
\end{enumerate}
\end{lemma}
\begin{IEEEproof}
 Notice that we are in Case 3: $1\leq
P_{L}\leq 66$, and that the state bits shift rightwards. So that
Lemma 11 is clear.
\end{IEEEproof}\vspace{0.2cm}

\begin{lemma}
Take the state at time 98 as the changed value as described in Lemma
11. Then for each $t$ such that $t\geq98$, each $j$ such that
$178\leq j\leq288$, $s_{(t+87,j)}=s_{(t,j)}+b_{t+265-j}.$
\end{lemma}
\begin{IEEEproof}
 The proof of Lemma 12 is somewhat
similar to that of Lemma 7. The proving details are the follow.
\begin{enumerate}
  \item If $178\leq j \leq 264$ and $t\geq 98$,then $t+265-j\geq 99$, so
  that
\begin{displaymath}
\left.
\begin{array}{ll}
s_{(t+87,j)}\hspace{-0.3cm}&=s_{(t+265-j,178)}\\
       &=s_{(t+264-j,264)}+a_{t+265-j}\\
       &=s_{(t,j)}+a_{t+265-j}.
  \end{array}\right.
  \end{displaymath}

  \item If $265\leq j \leq288$ and $t\geq 122$, then $241\leq j-24\leq264$ and $t-24\geq
 98$. By 1),
\begin{displaymath}
\left.
\begin{array}{ll}
s_{(t+87,j)}\hspace{-0.3cm}&=s_{(t-24+87,j-24)}\\
       &=s_{(t-24,j-24)}+a_{t-24+265-(j-24)}\\
       &=s_{(t,j)}+a_{t+265-j}.
  \end{array}\right.
  \end{displaymath}

  \item If $265\leq j \leq288$ and $t=98$, then $178\leq j-87\leq201$,
so $s_{(98+87,j)}=s_{(98,j-87)}$. By the assumptions of Lemma 11,
\begin{displaymath}
\left.
\begin{array}{ll}
s_{(98+87,j)}\hspace{-0.3cm}&=s_{(98,j-87)}\\
       &=s_{(98,j)}+a_{98+265-j}.
  \end{array}\right.
  \end{displaymath}

  \item If $265\leq j \leq288,99\leq t\leq121$, and $j-(t-98)\leq 264$,
then $242\leq j-(t-98)\leq 264$. By 1),
\begin{displaymath}
\left.
\begin{array}{ll}
s_{(t+87,j)}\hspace{-0.3cm}&=s_{(98+87,j-(t-98))}\\
       &=s_{(98,j-(t-98))}+a_{98+265-(j-(t-98))}\\
       &=s_{(t,j)}+a_{t+265-j}.
  \end{array}\right.
  \end{displaymath}

  \item If $265\leq j \leq288,99\leq t\leq121$, and $j-(t-98)\geq
265$, then $265\leq j-(t-98)\leq 287$. By 3),
\begin{displaymath}
\left.
\begin{array}{ll}
s_{(t+87,j)}\hspace{-0.3cm}&=s_{(98+87,j-(t-98))}\\
       &=s_{(98,j-(t-98))}+a_{98+265-(j-(t-98))}\\
       &=s_{(t,j)}+a_{t+265-j}.
  \end{array}\right.
  \end{displaymath}

\end{enumerate}
  Lemma 12 is proved.
  \end{IEEEproof}
\vspace{0.2cm}

\begin{lemma}
Take the state at time 98 as the changed value as
 described in Lemma 11. Then
\begin{enumerate}
  \item For each $t$ such that $t\geq98$, each $j$ such that $178\leq
j\leq 288$,
   $$s_{(t+2262,j)}=s_{(t,j)}+\sum_{m=0}^{25}b_{t+31-j+3m}.$$

  \item $\{(s_{(t,94)},\cdots,s_{(t,288)},t\geq98)\}$has a
period 4524.
\end{enumerate}
\end{lemma}
  \begin{IEEEproof}
According to Lemma $10$, Lemma $11$, Lemma $12$ and the fact that
$2262=87\times 26=78\times29$,
\begin{displaymath}
\left.
\begin{array}{ll}
s_{(t+2262,j)}\hspace{-0.3cm} \vspace{0.2cm}&=s_{(t+87\times26,j)}\\
     \vspace{0.2cm}  &=s_{(t,j)}+\sum_{n=0}^{25}b_{t+265-j+87\times n(mod 78)}\\
       &=s_{(t,j)}+\sum_{m=0}^{25}b_{t+31-j+3m},
  \end{array}\right.
  \end{displaymath}
  so that 1) is true. According to 1), for each $t$ such that $t\geq98$, each $j$ such that $178\leq
 j\leq 288$,
 \begin{displaymath}
 \left.
 \begin{array}{ll}
 s_{(t+4524,j)}\hspace{-0.3cm} \vspace{0.2cm}&=s_{(t+2262+2262,j)}\\
     \vspace{0.2cm}  &=s_{(t,j)}+\sum_{m=0}^{25}b_{t+31-j+3m}\\
          \vspace{0.2cm} &\hspace{0.4cm}+\sum_{m=0}^{25}b_{t+2262+31-j+3m}\\
       &=s_{(t,j)},
 \end{array}\right.
  \end{displaymath}
  This implies that $\{(s_{(t,178)},\cdots,s_{(t,288)},t\geq98)\}$ has a period $4524$.
  Again by the fact that $\{(s_{(t,94)},\cdots,s_{(t,177)},t\geq98)\}$ has a period $78$, Lemma 13 is proved.
    \end{IEEEproof}
\vspace{0.2cm}

\begin{proposition}
Suppose $1\leq P_{L}\leq66$. Then
\begin{enumerate}
  \item The key--stream $(z_{0}z_{1}z_{2}\cdots)$ has a period 4524.

  \item $(z_{0},z_{1},z_{2},\cdots,z_{4523})$ are linear functions
of $243$ variables
$(s_{(98,100)},\cdots,s_{(98,177)},s_{(98,202)},\cdots,
s_{(98,288)},\\
 b_{99},\cdots, b_{176})$, and these functions are
known.

  \item By knowing the values of $(z_{0},z_{1},z_{2},\cdots,z_{4523})$,
the attacker obtains $4524$ linear equations of $243$ variables
$(s_{(98,100)},\cdots,s_{(98,177)},s_{(98,202)},\cdots,s_{(98,288)},b_{99},\cdots,\\
b_{176})$. The rank of these linear equations is $237$, so that
there are $2^{6}=64$ possible solutions.
\end{enumerate}
\end{proposition}
  \begin{IEEEproof}
 1) is clear from Lemma 13. Notice that
for each $t$ such that $t\geq98$,

$(s_{(t+1,94)},\cdots,s_{(t+1,177)})=(s_{(t,171)},s_{(t,94)},\cdots,s_{(t,176)}),$

 \hspace{1.2cm}$(s_{(t+1,178)},\cdots,s_{(t+1,288)})$

 \hspace{0.8cm}$=(s_{(t,264)}+b_{t+1},s_{(t,178)},\cdots,s_{(t,287)}).$

So that, for each $t$ such that $t\geq98$,
$(s_{(t,94)},\cdots,s_{(t,288)})$\\
can be induced from
$(s_{(98,100)},\cdots,s_{(98,177)},s_{(98,202)},\cdots,\\
s_{(98,288)},b_{99},\cdots, b_{176})$  by linear recursion which is
already known. So that 2) is true.

3) is our checking result. Proposition 3 is proved.
  \end{IEEEproof}\vspace{0.2cm}

  Notice that the true value of $(s_{(98,100)},\cdots,s_{(98,177)},\\s_{(98,202)},\cdots,
  s_{(98,288)},b_{99},\cdots,b_{176})$ satisfies $78$ non--linear equations, described as the
follow.

$b_{99}=s_{(98,162)}+s_{(98,175)}s_{(98,176)}+s_{(98,177)},$

$b_{100}=s_{(98,161)}+s_{(98,174)}s_{(98,175)}+s_{(98,176)},$

$\cdots$

$b_{161}=s_{(98,100)}+s_{(98,113)}s_{(98,114)}+s_{(98,115)},$

$b_{162}=s_{(98,177)}+s_{(98,112)}s_{(98,113)}+s_{(98,114)},$

$b_{163}=s_{(98,176)}+s_{(98,111)}s_{(98,112)}+s_{(98,113)},$

$\cdots$

$b_{174}=s_{(98,165)}+s_{(98,100)}s_{(98,101)}+s_{(98,102)},$

$b_{175}=s_{(98,164)}+s_{(98,177)}s_{(98,100)}+s_{(98,101)},$

$b_{176}=s_{(98,163)}+s_{(98,176)}s_{(98,177)}+s_{(98,100)}.$

 78 non--linear equations and 4524 linear equations are\\
  enough to
 determine the true value of $(s_{(98,100)},\cdots,\\
 s_{(98,177)},b_{99},\cdots,b_{176})$.
 They are not enough to determine the true value of $(s_{(98,202)},\cdots,
  s_{(98,288)})$ because, in each linear equation, just 2
 variables of $(s_{(98,202)},\cdots,
  s_{(98,288)})$ appear. After that
 determination, 4524 linear equations become the linear equations of
 87 variables $(s_{(98,202)},\cdots,
  s_{(98,288)})$, and we have verified that
 the rank of these linear equations is 86. This fact restricts $(s_{(98,202)},\cdots,
  s_{(98,288)})$ into 2 possible values.

 Then we redefine
 $\{a_{t+1},t\geq0\}$. For each $t$ such that $t\geq0$, $a_{t+1}=s_{(t,66)}+s_{(t,91)}s_{(t,
 92)}+s_{(t, 93)}$. By considering Lemma 9, $a_{t+1}=0$ for each $t$ such that
 $t\geq92$.
\vspace{0.2cm}

\begin{lemma}{\ }
\begin{enumerate}
  \item $(s_{(98,94)},\cdots,s_{(98,177)})=(a_{20},a_{19},\cdots,a_{15},a_{14}+a_{92},$

\hspace{1cm}
$a_{13}+a_{91},\cdots,a_{1}+a_{79},a_{78},a_{77}\cdots,a_{15}).$

  \item $(s_{(98,178)},\cdots,s_{(98,288)})=(a_{29},a_{28},\cdots,a_{1},0,\cdots,0,$

\hspace{1cm} $b_{98}+a_{29},b_{97}+a_{28},\cdots,b_{75}+a_{6}).$
\end{enumerate}
\end{lemma}

(this is the changed value according to Lemma 11)
\begin{IEEEproof}
 We induce the state at time 98 by
gradually renewing the state.

\begin{enumerate}
  \item $(s_{(78,94)},\cdots,s_{(78,177)})=(a_{78},a_{77},\cdots,a_{1},0,\cdots,0),$
$(s_{(84,94)},\cdots,s_{(84,177)})=(a_{6}+a_{84},a_{5}+a_{83},\cdots,$

\hspace{3.5cm}$a_{1}+a_{79},a_{78},a_{77},\cdots,a_{1}),$

$(s_{(92,94)},\cdots,s_{(92,177)})=(a_{14}+a_{92},a_{13}+a_{91},\cdots,$

\hspace{3.3cm}$a_{1}+a_{79},a_{78},a_{77},\cdots,a_{9}),$

$(s_{(98,94)},\cdots,s_{(98,177)})=(a_{20},a_{19},\cdots,a_{15},a_{14}+a_{92},$

\hspace{1cm}$a_{13}+a_{91},\cdots,a_{1}+a_{79},a_{78},a_{77},\cdots,a_{15}).$

  \item $(s_{(69,178)},\cdots,s_{(69,288)})=(0,\cdots,0),$

$(s_{(78,178)},\cdots,s_{(78,288)})=(a_{9},a_{8},\cdots,a_{1},0,\cdots,0),$

$(s_{(98,178)},\cdots,s_{(98,288)})=(a_{29},a_{28},\cdots,a_{1},0,\cdots,0).$\\
But the value of ($(s_{(98,265)},\cdots,s_{(98,288)})$ is changed
according to Lemma 11, so that\\
$(s_{(98,178)},\cdots,s_{(98,288)})=(a_{29},a_{28},\cdots,a_{1},0,\cdots,0,$

\hspace{1.5cm} $b_{98}+a_{29},b_{97}+a_{28},\cdots,b_{75}+a_{6}).$
\end{enumerate}
 Lemma 14
is proved.
\end{IEEEproof} \vspace{0.2cm}

 Lemma 14 shows $(s_{(98,207)},s_{(98,208)},\cdots,s_{(98,264)})=(0,\cdots,0)$.
This fact and all former equations are enough to determine the true
value of $(s_{(98,202)},\cdots,s_{(98,288)})$.

 Up to now, 243
variables $\{s_{(98,100)},\cdots,s_{(98,177)},s_{(98,202)},\\
\cdots,s_{(98,288)},b_{99},\cdots,b_{176}\}$ have already been
uniquely determined. According to Lemma 14, the attacker can solve
the value of $(a_{1},a_{2},\cdots,a_{92})$, which is the closest to
the key $(k_{1},\cdots,k_{80})$. $(a_{1},a_{2},\cdots,a_{92})$ is an
unknown function of $(k_{1},\cdots,k_{80})$, because hard fault
positions are unknown. But $(a_{1},a_{2},\cdots,a_{92})$ can
partially reveal the key, as described in Proposition 4 and
Proposition 5. \vspace{0.2cm}

\begin{lemma}
Suppose the indices of hard--fault--injected--bits are not from the
set $\{j,j+1,\cdots,j+m\}$, where $1\leq j\leq j+m\leq93$. Then
$s_{(m,j+m)}=s_{(0,j)}$.
\end{lemma}\vspace{0.2cm}

\begin{proposition}
Suppose $1\leq P_{L}\leq66$. Suppose $a_{t+1}=1$ for some $t$ such
that $0\leq t\leq11$. Then

\hspace{0.6cm}$(a_{1},a_{2},\cdots,a_{t+1})=(k_{66},k_{65},\cdots,k_{66-t}).$
\end{proposition}
\begin{IEEEproof}
 Notice that

\hspace{1cm}$(s_{(0,81)},s_{(0,82)},\cdots,s_{(0,93)})=(0,\cdots,0),$\\
 so that

 $(s_{(0,91)}s_{(0,92)}+s_{(0,93)},s_{(1,91)}s_{(1,92)+s(1,93)},\cdots,$

\hspace{0.7cm} $s_{(12,91)}s_{(12,92)}+s_{(12,93)})=(0,\cdots,0),$\\
  and that

$(a_{1},a_{2},\cdots,a_{12})=(s_{(0,66)},s_{(1,66)},\cdots,s_{(12,66)}).$\\
Suppose $a_{t+1}=1$ for some $t$ such that $0\leq t\leq11$, then the
indices of hard--fault--injected--bits are never from the set
$\{66-t, 67-t,\cdots,66\}$, or else there would be a contradiction.
According to Lemma 15,
\begin{displaymath}
 \left.
 \begin{array}{ll}
 (a_{1},a_{2},\cdots,a_{t+1})\hspace{-0.3cm} &=(s_{(0,66)},s_{(1,66)},\cdots,s_{(t,66)})\\
     &=(s_{(0,66)},s_{(0,65)},\cdots,s_{(0,66-t)})\\
       &=(k_{66},k_{65},\cdots,k_{66-t}).
 \end{array}\right.
  \end{displaymath}
 Proposition 4 is proved.
 \end{IEEEproof}
\vspace{0.2cm}

\begin{proposition}
Suppose $1\leq P_{L}\leq66$. Suppose $a_{t+1}=1$ for some $t$ such
that $67\leq t\leq91$. Then
\begin{enumerate}
  \item $(a_{1},a_{2},\cdots,a_{12})=(k_{66},k_{65},\cdots,k_{55}).$
  \item $a_{13}=k_{54}+k_{79}k_{80}.$

  \item Either a) or b) is true,
where
\begin{enumerate}
  \item $a_{u+1}=k_{66-u}+k_{91-u}k_{92-u}+k_{93-u}$ for
 $13\leq u\leq t-27$, and $a_{v+1}=k_{91-v}k_{92-v}+k_{93-v}$
for  $65\leq v\leq t-2$.
  \item $a_{u+1}=k_{66-u}+k_{91-u}k_{92-u}$ for $13\leq
u\leq t-27$, and $a_{v+1}=k_{91-v}k_{92-v}$ for $65\leq v\leq t-2$.
\end{enumerate}
\end{enumerate}
\end{proposition}
\begin{IEEEproof}
By the assumption $"1\leq P_{L}\leq66"$ we know that
$(s_{(65,66)},s_{(66,66)},\cdots,s_{(91,66)})=(0,\cdots,0)$, so that

 $(a_{66},a_{67},\cdots,a_{92})\hspace{-0.1cm}
 =(s_{(65,91)}s_{(65,92)}+s_{(65,93)},$

 $s_{(66,91)}s_{(66,92)}+s_{(66,93)},\cdots,s_{(91,91)}s_{(91,92)}+s_{(91,93)}).$

  Suppose $a_{t+1}=s_{(t,91)}s_{(t,92)}+s_{(t,93)}=1$ for some $t$ such that $67\leq t\leq91$, then
  the indices of hard--fault--injected--bit are never from the set
$\{93-t,94-t,\cdots,92\}$, or else there would be a contradiction.
Notice that
$(a_{1},a_{2},\cdots,a_{12})=(s_{(0,66)},s_{(1,66)},\cdots,s_{(11,66)})$.
So that
\begin{displaymath}
 \left.
 \begin{array}{ll}
 (a_{1},a_{2},\cdots,a_{12})\hspace{-0.3cm} &=(s_{(0,66)},s_{(1,66)},\cdots,s_{(11,66)})\\
      &=(s_{(0,66)},s_{(0,65)},\cdots,s_{(0,55)})\\
           &=(k_{66},k_{65},\cdots,k_{55}).
 \end{array}\right.
  \end{displaymath}
 \begin{displaymath}
 \left.
 \begin{array}{ll}
a_{13}\hspace{-0.3cm}&=s_{(12,66)}+s_{(12,91)}s_{(12,92)}+s_{(12,93)}\\
      &=s_{(0,54)}+s_{(0,79)}s_{(0,80)}+s_{(0,81)}\\
           &=k_{54}+k_{79}k_{80}.
 \end{array}\right.
  \end{displaymath}
 1) and 2) are true.

 Now suppose that 93 is not an index of hard--fault--injected--bit.

For each $u$ such that $13\leq u\leq t-27$, we have $93-t\leq
66-u<91-u<92-u<93-u\leq80$, so that
\begin{displaymath}
 \left.
 \begin{array}{ll}
a_{u+1}\hspace{-0.3cm}&=s_{(u,66)}+s_{(u,91)}s_{(u,92)}+s_{(u,93)}\\
       &=s_{(0,66-u)}+s_{(0,91-u)}s_{(0,92-u)}+s_{(0,93-u)}\\
           &=k_{66-u}+k_{91-u}k_{92-u}+k_{93-u}.
 \end{array}\right.
  \end{displaymath}
For each $v$ such that $65\leq v\leq t-2$, we have $93-t\leq
91-v<92-v<93-v\leq28$, so that
\begin{displaymath}
 \left.
 \begin{array}{ll}
a_{v+1}\hspace{-0.3cm} &=s_{(v,66)}+s_{(v,91)}s_{(v,92)}+s_{(v,93)}\\
      &=s_{(v,91)}s_{(v,92)}+s_{(v,93)}\\
      &=s_{(0,91-v)}s_{(0,92-v)}+s_{(0,93-v)}\\
          &=k_{91-v}k_{92-v}+k_{93-v}.
 \end{array}\right.
  \end{displaymath}
  a) is true.

  Now suppose that 93 is an index of hard--fault--injected--bit. Then
  $s_{(0,93)}=s_{(1,93)}=\cdots=s_{(91,93)}=0.$

For each $u$ such that $13\leq u\leq t-27$, we have $93-t\leq
66-u<91-u<92-u\leq79$, so that
\begin{displaymath}
 \left.
 \begin{array}{ll}
a_{u+1}\hspace{-0.3cm} &=s_{(u,66)}+s_{(u,91)}s_{(u,92)}+s_{(u,93)}\\
       &=s_{(u,66)}+s_{(u,91)}s_{(u,92)}\\
       &=s_{(0,66-u)}+s_{(0,91-u)}s_{(0,92-u)}\\
           &=k_{66-u}+k_{91-u}k_{92-u}.
 \end{array}\right.
  \end{displaymath}
  For each $v$ such that $65\leq v\leq t-2$, we have $93-t\leq
91-v<92-v\leq27$, so that
\begin{displaymath}
 \left.
 \begin{array}{ll}
a_{v+1}\hspace{-0.3cm} &=s_{(v,66)}+s_{(v,91)}s_{(v,92)}+s_{(v,93)}\\
     &=s_{(v,91)}s_{(v,92)}\\
     &=s_{(0,91-v)}s_{(0,92-v)}\\
           &=k_{91-v}k_{92-v}.
 \end{array}\right.
  \end{displaymath}
  Proposition 5 is proved.
  \end{IEEEproof}
  \subsection{Features of Fault Injected Machine in Case 4:
  $163\leq P_{L}\leq 171$}
\begin{proposition}
Suppose we are in Case 4: $163\leq P_{L}\leq 171$. Then
\begin{enumerate}
  \item For each $t$ such that $t\geq 0$,

\hspace{1.5cm}$(s_{(t,171)},\cdots,s_{(t,177)})=(0,\cdots,0)$,\\
so that generation of the key--stream $(z_{0}z_{1}z_{2}\cdots)$ is
degraded as
\begin{displaymath}
 \left.
 \begin{array}{ll}
z_{t}\hspace{-0.3cm} &=s_{(t+1152,66)}+s_{(t+1152,93)}\\
     &+s_{(t+1152,162)}+s_{(t+1152,243)}+s_{(t+1152,288)},t\geq 0.
 \end{array}\right.
  \end{displaymath}
and the state is degraded into 273 bits\\
$(s_{(t,1)},s_{(t,2)},\cdots,s_{(t,162)},s_{(t,178)},s_{(t,179)},\cdots,
s_{(t,288)})$.

  \item The state renewal is the follow.

   \hspace{0.5cm}$(s_{(t+1,1)},s_{(t+1,2)},\cdots,s_{(t+1,93)})$

  \hspace{0.2cm}$=(s_{(t,243)}+s_{(t,286)}s_{(t,287)}+s_{(t,288)}+s_{(t,69)},$

  \hspace{0.5cm}$s_{(t,1)},\cdots,s_{(t,92)}),$

  \hspace{0.5cm}$(s_{(t+1,94)},s_{(t+1,95)},\cdots,s_{(t+1,162)})$

  \hspace{0.2cm}$=(s_{(t,66)}+s_{(t,91)}s_{(t,92)}+s_{(t,93)},s_{(t,94)},\cdots,s_{(t,161)}),$

   \hspace{0.5cm}$(s_{(t+1,178)},s_{(t+1,179)},\cdots,s_{(t+1,288)})$

  \hspace{0.2cm}$=(s_{(t,162)}+s_{(t,264)},s_{(t,178)},\cdots,s_{(t,287)}).$

\item  The state renewal is reversible, and the inverse is the follow.

      \hspace{0.5cm}$(s_{(t,1)},s_{(t,2)},\cdots,s_{(t,93)})$

   \hspace{0.2cm}$=(s_{(t+1,2)},s_{(t+1,3)},\cdots,s_{(t+1,93)},$

  \hspace{0.6cm}
   $s_{(t+1,67)}+s_{(t+1,92)}s_{(t+1,93)}+s_{(t+1,94)}),$

   \hspace{0.5cm}$(s_{(t,94)},s_{(t,95)},\cdots,s_{(t,162)})$

   \hspace{0.2cm}$=(s_{(t+1,95)},s_{(t+1,96)},\cdots,s_{(t+1,162)},$

  \hspace{0.6cm}$s_{(t+1,178)}+s_{(t+1,265)}),$

 \hspace{0.5cm}$(s_{(t,178)},s_{(t,179)},\cdots,s_{(t,288)})$

 \hspace{0.2cm}$=(s_{(t+1,179)},s_{(t+1,180)},\cdots,s_{(t+1,288)},$

\hspace{0.4cm}
$s_{(t+1,244)}+s_{(t+1,287)}s_{(t+1,288)}+s_{(t+1,1)}+s_{(t+1,70)}).$

  \item Change the IV (Initial Vector) from
$(IV_{1},\cdots,IV_{80})=(0,\cdots,0)$ to the follow: $IV_{j}=0$ for
each $j$ such that $1\leq j\leq80$, except $IV_{70}=1$. Then the
key--stream $(z_{0}z_{1}z_{2}\cdots)$ are kept unchanged.
\end{enumerate}
\end{proposition}\vspace{0.2cm}

  Proposition 6 is clear by considering Trivium key--stream generation
and Trivium state renewal. The following Proposition 7 is our
checking result.\vspace{0.2cm}

\begin{proposition}
Suppose we are in Case 4: $163\leq P_{L}\leq171$. Let
$(s_{1},\cdots,s_{162},s_{178},\cdots,s_{288})$ denote the initial
state (that is, the state at the time just before generating
$z_{0}$). Take $\{z_{0},z_{1},z_{2},\cdots\}$ as functions of
$(s_{1},\cdots,s_{162},s_{178},\cdots,s_{288})$. Then
\begin{enumerate}
  \item $\{z_{0},z_{1},\cdots,z_{65}\}$ are 66 linear functions.
  \item $\{z_{66},z_{67},\cdots,z_{159}\}$ are 94 quadratic functions.
  \item $\{z_{160},z_{161},\cdots,z_{228}\}$ are 69 cubic functions.
  \item Each of $\{z_{229},z_{230},\cdots\}$ is at least a quartic function.
\end{enumerate}
\end{proposition}\vspace{0.2cm}

 Proposition 6 and Proposition 7 present a simpler cipher than
Trivium. It has a smaller number of state bits and a slower
non--linearization procedure. So that it is easier to solve the
state at a fixed time. If the state at a fixed time is known, the
key will be known by reversing the state.
  \subsection{Features of Fault Injected Machine in Case 5:
  $172\leq P_{L}\leq 176$}
\begin{lemma}
Suppose we are in Case 5: $172\leq
P_{L}\leq 176$. Then
\begin{enumerate}
  \item For each $t$ such that $t\geq 5$,

\hspace{1.5cm}$(s_{(t,176)},s_{(t,177)})=(0,0)$.

  \item Suppose $m$ is the earliest time such that, for
  each $t\geq m$, $(s_{(t,176)},s_{(t,177)})=(0,0)$. Then for each $t\geq m$, we have
  \begin{enumerate}
    \item The state is degraded into 282 bits\\
    $(s_{(t,1)},s_{(t,2)},\cdots,s_{(t,171)},s_{(t,178)},s_{(t,179)},\cdots,\\
    s_{(t,288)})$.
    \item State renewal is the follow.

 \hspace{0.3cm}$(s_{(t+1,1)},s_{(t+1,2)},\cdots,s_{(t+1,93)})$

$=(s_{(t,243)}+s_{(t,286)}s_{(t,287)}+s_{(t,288)}+s_{(t,69)},$

 \hspace{0.3cm}$s_{(t,1)},\cdots,s_{(t,92)}),$

 \hspace{0.3cm}$(s_{(t+1,94)},s_{(t+1,95)},\cdots,s_{(t+1,171)})$

 $=(s_{(t,66)}+s_{(t,91)}s_{(t,92)}+s_{(t,93)}+s_{(t,171)},$

       \hspace{0.3cm} $s_{(t,94)},\cdots,s_{(t,170)}),$

 \hspace{0.3cm}$(s_{(t+1,178)},s_{(t+1,179)},\cdots,s_{(t+1,288)})$

  $=s_{(t,162)}+s_{(t,264)},s_{(t,178)},\cdots,s_{(t,287)}).$
  \end{enumerate}
\end{enumerate}
\end{lemma}\vspace{0.2cm}

Lemma 16 is clear by considering Trivium key--stream generation and
Trivium state renewal. Notice that state renewal procedure in Lemma
16-2)-b) is irreversible. \vspace{0.2cm}

\begin{lemma}
Suppose $m$ is the earliest time
such that, for each $t\geq m$, $(s_{(t,176)},s_{(t,177)})=(0,0)$.
Then
\begin{description}
  \item[1)]  \hspace{-0.5cm}For each $t$ such that $t\geq m+1$,

$s_{(t,163)}+s_{(t,178)}+s_{(t,265)}=0.$
  \item[2)]\hspace{-0.5cm}For each $t$ such that $t\geq m+2$,

$s_{(t,164)}+s_{(t,179)}+s_{(t,266)}=0.$

  \item[] \hspace{-0.5cm}$\cdots$
  \item [9)] \hspace{-0.5cm}For each $t$ such that $t\geq m+9$,

 $s_{(t,171)}+s_{(t,186)}+s_{(t,273)}=0.$
\end{description}
\end{lemma}
\begin{IEEEproof}
By Lemma 16 we know that, for each $t$ such that $t\geq m+1$,

 \hspace{1.5cm} $s_{(t,163)}=s_{(t-1,162)},$

 \hspace{1.5cm} $s_{(t,178)}=s_{(t-1,162)}+s_{(t-1,264)},$

 \hspace{1.5cm} $s_{(t,265)}=s_{(t-1,264).}$

So that 1) is true. Again for each $t$ such that $t\geq m+1$,
\begin{displaymath}
 \left.
 \begin{array}{ll}
&\hspace{0.3cm}s_{(t,163)}+s_{(t,178)}+s_{(t,265)}\\
  &=s_{(t+1,164)}+s_{(t+1,179)}+s_{(t+1,266)}\\
     & \hspace{0.3cm}\cdots\\
     &=s_{(t+8,171)}+s_{(t+8,186)}+s_{(t+8,273)}.
 \end{array}\right.
  \end{displaymath}

  So that $2),3),\cdots,9)$ are true, by considering 1). Lemma 17 is proved.
  \end{IEEEproof}
\vspace{0.2cm}

 \begin{proposition}
 Suppose we are in Case 5: $172\leq P_{L}\leq176$. Then
\begin{enumerate}
  \item Generation of the key--stream $(z_{0}z_{1}z_{2}\cdots)$ is
degraded as
\begin{displaymath}
 \left.
 \begin{array}{ll}
z_{t}\hspace{-0.3cm} &=s_{(t+1152,66)}+s_{(t+1152,93)}\\
     &+s_{(t+1152,162)}+s_{(t+1152,243)}+s_{(t+1152,288)},t\geq 0.
 \end{array}\right.
  \end{displaymath}

  \item Suppose $m$ is the earliest time such that, for each $t\geq m$, $(s_{(t,176)},s_{(t,177)})=(0,0)$.
  Then for each $t\geq m+9$, we have
       \begin{enumerate}
         \item the state is degraded into 273 bits\\
$(s_{(t,1)},s_{(t,2)},\cdots,s_{(t,162)},s_{(t,178)},s_{(t,179)},\cdots,\\
s_{(t,288)})$.

         \item The state renewal is the follow.

 \hspace{0.4cm}$(s_{(t+1,1)},s_{(t+1,2)},\cdots,s_{(t+1,93)})$

$=(s_{(t,243)}+s_{(t,286)}s_{(t,287)}+s_{(t,288)}+s_{(t,69)},$

 \hspace{0.6cm}$s_{(t,1)},\cdots,s_{(t,92)}),$

 \hspace{0.4cm}$(s_{(t+1,94)},s_{(t+1,95)},\cdots,s_{(t+1,162)})$

 $=(s_{(t,66)}+s_{(t,91)}s_{(t,92)}+s_{(t,93)}+s_{(t,186)}+s_{(t,273)},$

 \hspace{0.6cm}$s_{(t,94)},\cdots,s_{(t,161)}),$

 \hspace{0.4cm}$(s_{(t+1,178)},s_{(t+1,179)},\cdots,s_{(t+1,288)})$

$=(s_{(t,162)}+s_{(t,264)},s_{(t,178)},\cdots,s_{(t,287)}).$

         \item The state renewal is reversible, and the inverse is the
follow.

 \hspace{0.4cm}$(s_{(t,1)},s_{(t,2)},\cdots,s_{(t,93)})$

$=(s_{(t+1,2)},s_{(t+1,3)},\cdots,s_{(t+1,93)},$

\hspace{0.4cm}$s_{(t+1,67)}+s_{(t+1,92)}s_{(t+1,93)}+s_{(t+1,94)}+$

\hspace{0.4cm}$s_{(t+1,187)}+s_{(t+1,274)}),$

\hspace{0.4cm}$(s_{(t,94)},s_{(t,95)},\cdots,s_{(t,162)})$

$=(s_{(t+1,95)},s_{(t+1,96)},\cdots,s_{(t+1,162)},$

\hspace{0.4cm}$s_{(t+1,178)}+s_{(t+1,265)}),$

 \hspace{0.4cm}$(s_{(t,178)},s_{(t,179)},\cdots,s_{(t,288)})$

$=(s_{(t+1,179)},s_{(t+1,180)},\cdots,s_{(t+1,288)},s_{(t+1,1)}+$

 \hspace{0.4cm}
$s_{(t+1,70)}+s_{(t+1,244)}+s_{(t+1,287)}s_{(t+1,288)}).$

       \end{enumerate}
  \item Change the IV (Initial Vector) from
$(IV_{1},\cdots,IV_{80})=(0,\cdots,0)$ to the follow: $IV_{j}=0$ for
each $j$ such that $1\leq j\leq80$, except $IV_{79}=1$. Then the
key--stream $(z_{0}z_{1}z_{2}\cdots)$ are kept unchanged.
\end{enumerate}
 \end{proposition}
\begin{IEEEproof}
 1) is clear. 2) is a natural corollary of Lemma 16
and Lemma 17. 3) is clear.
 \end{IEEEproof}\vspace{0.2cm}

The following Proposition 9 is our checking result. \vspace{0.2cm}

 \begin{proposition}
Suppose we are in Case 5: $172\leq P_{L}\leq176$. Let
$(s_{1},\cdots,s_{162},s_{178},\cdots,s_{288})$ denote the initial
state (that is, the state at the time just before generating
$z_{0}$). Take $\{z_{0},z_{1},z_{2},\cdots\}$ as functions of
$(s_{1},\cdots,s_{162},s_{178},\cdots,s_{288})$. Then
\begin{enumerate}
  \item $\{z_{0},z_{1},\cdots,z_{65}\}$ are 66 linear functions.
  \item $\{z_{66},z_{67},\cdots,z_{159}\}$ are 94 quadratic functions.
  \item $\{z_{160},z_{161},\cdots,z_{228}\}$ are 69 cubic functions.
  \item Each of $\{z_{229},z_{230},\cdots\}$is at least a quartic function.
\end{enumerate}
 \end{proposition}\vspace{0.2cm}

  Proposition 8 and Proposition 9 present a simpler cipher than
Trivium. It has a smaller number of state bits and a slower
non--linearization procedure. So that it is easier to solve the
state at a fixed time. If the state at a fixed time is known, the
state at time 14 will be known by reversing the state, described in
Proposition 8 (we know that $14\geq m+9$, where $m$ is the earliest
time such that, for each $t\geq m$, $(s_{(t,176)},s_{(t,177)})=(0,
0)$).

   Now suppose that the state at time 14 is known. We know that
$(k_{1},\cdots,k_{79})=(s_{(14,15)},s_{(14,16)},\cdots,s_{(14,93)})$.
Then, if
$m<5,k_{80}=s_{(13,93)}=s_{(14,67)}+s_{(14,92)}s_{(14,93)}+s_{(14,94)}+s_{(14,187)}+s_{(14,274)}$,
according to Proposition 8. If $m=5$, the value of $k_{80}$ can not
be determined.
 \subsection{Features of Fault Injected Machine in Case 6:
  $P_{L}=177$}
\begin{proposition}
Suppose we are in Case 6: $ P_{L}=177$. Then
\begin{enumerate}
  \item Generation of the key--stream $(z_{0}z_{1}z_{2}\cdots)$ is degraded
as
\begin{displaymath}
 \left.
 \begin{array}{ll}
z_{t}\hspace{-0.3cm} &=s_{(t+1152,66)}+s_{(t+1152,93)}\\
     &+s_{(t+1152,162)}+s_{(t+1152,243)}+s_{(t+1152,288)},t\geq 0.
 \end{array}\right.
  \end{displaymath}

  \item the state is degraded into 287 bits\\
$(s_{(t,1)},s_{(t,2)},\cdots,s_{(t,176)},s_{(t,178)},s_{(t,179)},\cdots,
s_{(t,288)})$.
  \item The state renewal is the follow.

 \hspace{0.4cm}$(s_{(t+1,1)},s_{(t+1,2)},\cdots,s_{(t+1,93)})$

$=(s_{(t,243)}+s_{(t,286)}s_{(t,287)}+s_{(t,288)}+s_{(t,69)},$

 \hspace{0.5cm}$s_{(t,1)},\cdots,s_{(t,92)}),$

 \hspace{0.4cm}$(s_{(t+1,94)},s_{(t+1,95)},\cdots,s_{(t+1,176)})$

  $=(s_{(t,66)}+s_{(t,91)}s_{(t,92)}+s_{(t,93)}+s_{(t,171)},$

  \hspace{0.5cm}$s_{(t,94)},\cdots,s_{(t,175)}),$

 \hspace{0.4cm}$(s_{(t+1,178)},s_{(t+1,179)},\cdots,s_{(t+1,288)})$

$=(s_{(t,162)}+s_{(t,175)}s_{(t,176)}+s_{(t,264)},$

     \hspace{0.5cm} $s_{(t,178)},\cdots,s_{(t,287)}).$

  \item  Change the IV (Initial Vector) as
$(IV_{1},\cdots,IV_{78})=(0,\cdots,0)$, and $(IV_{79},IV_{80})\neq
(0,0)$. Then the key--stream $(z_{0}z_{1}z_{2}\cdots)$ are kept
unchanged.
\end{enumerate}
\end{proposition}\vspace{0.2cm}

Proposition 10 is clear. Notice that state renewal is irreversible.
\subsection{Features of Fault Injected Machine in Case 7:
  $67\leq P_{L}\leq 93$ or $244\leq P_{L}\leq 288$}
Case 7 has many features similar with former cases. Here are some
examples.

If $244\leq P_{L}\leq264$, the features are similar to those of Case
4.

If $265\leq P_{L}\leq287$, the features are similar to those of Case
5.

If $ P_{L}=288$, the features are similar to those of Case 6.

If $67\leq P_{L}\leq69$, the features are similar to those of \\Case
4.

If $70\leq P_{L}\leq92$, the features are similar to those of \\Case
5.

If $P_{L}=93$, the features are similar to those of Case 6.

\section{Cases Checking}
In this section we present an algorithm, to check the case by
observing the key--stream $(z_{0}z_{1}z_{2}\cdots)$. We firstly
define 6 features for $(z_{0}z_{1}z_{2}\cdots)$.

 Feature
1:\hspace{0.2cm}$(z_{0}z_{1}\cdots,z_{68})=(z_{69}z_{70}\cdots
z_{137}).$

Feature 2:\hspace{0.2cm}
$(z_{0}z_{1}\cdots,z_{3357})=(z_{3358}z_{3359}\cdots z_{6715}).$

Feature
3:\hspace{0.2cm}$(z_{0}z_{1}\cdots,z_{4523})=(z_{4524}z_{4525}\cdots
z_{9047}).$

Feature 4:\hspace{0.2cm} Change $IV_{70}$ from 0 to 1, then
$(z_{0}z_{1}z_{2}\cdots z_{287})$ are kept unchanged.

Feature 5:\hspace{0.2cm}Change $IV_{79}$ from 0 to 1, then\\
$(z_{0}z_{1}z_{2}\cdots z_{287})$are kept unchanged.

 Feature 6:
Change $IV_{80}$ from 0 to 1, then\\ $(z_{0}z_{1}z_{2}\cdots
z_{287})$are kept unchanged. \vspace{0.2cm}

Then we point out some facts, as the follow.
\begin{enumerate}
  \item In Case 1, $(z_{0}z_{1}z_{2}\cdots)$ satisfies Feature 1.
  \item In Case 2, $(z_{0}z_{1}z_{2}\cdots)$ satisfies Feature 2.
  \item In Case 3, $(z_{0}z_{1}z_{2}\cdots)$ satisfies Feature 3.
  \item In Case 4, $(z_{0}z_{1}z_{2}\cdots)$ satisfies Feature 4.
  \item In Case 5, $(z_{0}z_{1}z_{2}\cdots)$ satisfies Feature 5.
  \item In Case 5, $(z_{0}z_{1}z_{2}\cdots)$ may or may not satisfy
Feature 6.
  \item In Case 6, $(z_{0}z_{1}z_{2}\cdots)$ satisfies both Feature 5
and Feature 6.
\end{enumerate}

Then we present some natural assumptions, described in the follow.
\begin{enumerate}
  \item If the case is not Case 1, $(z_{0}z_{1}z_{2}\cdots)$ satisfies
Feature 1 with a neglectable probability.

  \item If the case is neither Case 1 nor Case 2,
$(z_{0}z_{1}z_{2}\cdots)$ satisfies Feature 2 with a neglectable
probability.

  \item If the case is not from {Case 1, Case 2, Case
3},\\$(z_{0}z_{1}z_{2}\cdots)$ satisfies Feature 3 with a
neglectable probability.

  \item If the case is not from {Case 1, Case 2, Case 3, Case 4},\\
$(z_{0}z_{1}z_{2}\cdots)$ satisfies Feature 4 with a neglectable
probability.

  \item In Case 7, $(z_{0}z_{1}z_{2}\cdots)$ satisfies Feature 5 with a
neglectable probability.

  \item In Case 7, $(z_{0}z_{1}z_{2}\cdots)$ satisfies Feature 6 with a
neglectable probability.
\end{enumerate}
\vspace{0.2cm}

\emph{Algorithm} \hspace{0.1cm} Suppose that the attacker has
obtained the key--stream $(z_{0}z_{1}z_{2}\cdots)$, from a
hard--fault--injected machine.

\begin{enumerate}
  \item If $(z_{0}z_{1}z_{2}\cdots)$ satisfies Feature 1, take the case
as\\ Case 1.

  \item If $(z_{0}z_{1}z_{2}\cdots)$ does not satisfy Feature 1, but
satisfies Feature 2, take the case as Case 2.

  \item If $(z_{0}z_{1}z_{2}\cdots)$ does not satisfy each from {Feature
1, Feature 2}, but satisfies Feature 3, take the case as\\ Case 3.

  \item If $(z_{0}z_{1}z_{2}\cdots)$ does not satisfy each from {Feature
1, Feature 2, Feature 3}, but satisfies Feature 4, take the case as
Case 4.

  \item If $(z_{0}z_{1}z_{2}\cdots)$ does not satisfy each from {Feature
1, Feature 2, Feature 3, Feature 4}, but satisfies both Feature 5
and Feature 6, take the case as from {Case 5, Case 6}.

  \item If $(z_{0}z_{1}z_{2}\cdots)$ does not satisfy each from {Feature
1, Feature 2, Feature 3, Feature 4, Feature 6}, but satisfies
Feature 5, take the case as Case 5.

  \item If $(z_{0}z_{1}z_{2}\cdots)$ does not satisfy each from
{Feature 1, Feature 2, Feature 3, Feature 4, Feature 5, Feature 6},
take the case as Case 7.
\end{enumerate}

Under our natural assumptions, Algorithm selectes wrong cases with a
neglectable probability. In step 5) of Algorithm, we can also take
the case directly as Case 5. The probability of mistake is no more
than 1/5.
\section{Conclusion and Future Work}
From all of the discussions above, it is clear that Trivium is weak
under hard fault analysis, with our trivial assumptions.

Hard fault injection will lead us to continue our work. One future
work is combined fault analysis of Grain. Grain is another
hardware--oriented stream cipher, and one of the finally chosen
ciphers by eSTREAM project. We find Grain much stronger under either
soft or hard fault analysis. We will combine hard fault injection
and soft fault injection, looking for weakness of Grain. The second
future work is the study under weaker assumptions. One weaker
assumption is that, after fault injection, the values of those
injected bits are permanently 0 or 1.
\end{document}